\documentstyle[preprint,aps]{revtex}

\newcommand{\cH}{{\mathcal H}}
\newcommand{\cN}{{\mathcal N}}
\newcommand{\cP}{{\mathcal P}}
\newcommand{\tD}{\tilde{D}}
\newcommand{\tP}{\tilde{P}}

\newcommand{\tW}{\tilde{W}}

\tighten

\begin{document}

\draft

\title{General Forms of a $\cN$-fold Supersymmetric Family}

\author{Hideaki Aoyama$^{1}$\thanks{aoyama@phys.h.kyoto-u.ac.jp},
 Masatoshi Sato$^{2}$\thanks{msato@issp.u-tokyo.ac.jp} and
 Toshiaki Tanaka$^{1}$\thanks{ttanaka@phys.h.kyoto-u.ac.jp}}

\address{$^{1}$Faculty of Integrated Human Studies,\\
 Kyoto University, Kyoto 606-8501, Japan\\
$^{2}$The Institute for Solid State Physics,\\
 The University of Tokyo, Kashiwanoha 5-1-5,\\
 Kashiwa-shi, Chiba 277-8581, Japan}



\date{\today}

\maketitle

\begin{abstract}
We report general forms of one family of the $\cN$-fold
supersymmetry in one-dimensional quantum mechanics.
The $\cN$-fold supersymmetry is characterized by the supercharges
which are $\cN$-th order in differential operators. The family
reported here is defined as a particular form of the supercharges
and is referred to as ``type A''. We show that a quartic and
a periodic potentials, which were previously found to be $\cN$-fold
supersymmetric by the authors, are realized as special cases of
this type A family.
\end{abstract}




\section{Introduction}
\label{sec:intro}
Recently, much attention have been paid to the
{\it $\cN$-fold supersymmetry\/} in quantum mechanics as
one of the most fruitful generalization of the supersymmetry
\cite{And1,And2,And3,And4,Sam1,Sam2,Sam3,Sam4,Ply1,Ply2,Aoy1,Aoy2}. 
The $\cN$-fold supersymmetry
is characterized by a non-linear superalgebra among the supercharges
and the Hamiltonian; the anticommutator of the supercharges is
a polynomial of the Hamiltonian.
The coordinate representation of the supercharges
involves $\cN$-th order derivative.

There are several ways to construct the $\cN$-fold supersymmetric
models. If one has a Hamiltonian for which the exact $\cN$
eigenfunctions are known, the $\cN$-th order supercharge is given
by the $\cN$-th order Darboux transformation and can be represented
in the form known as Crum-Krein formula \cite{Sam1,Sam2}.
The formal expressions for the partner Hamiltonian and the
anticommutator of the supercharge are also known \cite{Sam1,Sam2}.
However, the applicability
of this approach will be, in practice, quite limited since we rarely
have exact solutions of a Hamiltonian under consideration. In addition,
serious difficulties will be expected when one intends to construct
a model in which dynamical SUSY breaking takes place. In this case,
the prepotential is related to (the logarithmic derivative of) the
{\it perturbative\/} ground-state eigenfunction which can be solved
analytically, but not to the {\it exact\/} ground-state eigenfunction.
Actually, we have already known a $\cN$-fold supersymmetric model
in which the purely nonperturbative effect breaks the $\cN$-fold
supersymmetry and only the perturbative $\cN$ non-degenerate states
can be obtained analytically \cite{Aoy1}.

Two $\cN$-fold supersymmetric models reported in Ref.\cite{Aoy1} and
Ref.\cite{Aoy2}
have the common significant features. The one is the simplicity of
the form of the potentials; in spite of the fact that the higher
order Darboux transformations generally lead to a quite complicated
form of the partner potential \cite{Sam3,Ros1,Kha1,Fer1,Fer2}.
The other is that the $\cN$-fold
supersymmetry for any $\cN$ are realized only through the specific
values of a parameter, say $\epsilon =\cN$, involved in {\it one\/}
Hamiltonian. These examples show the existence of $\cN$-fold
supersymmetric {\it family} in which a pair of the specific Hamiltonians
possesses any $\cN$-fold supersymmetry via one (or more) parameter(s)
involved in the Hamiltonians.

In this letter, we report the general forms and conditions of a system
to be $\cN$-fold supersymmetric family with respect to a particular
form of the supercharges, without any recourse to the information on
eigenfunctions. In section \ref{sec:nsusy}, we review the $\cN$-fold
supersymmetry including ordinary one. In section \ref{sec:typeA}, we
define a particular class of the $\cN$-fold supersymmetry, which will
be referred to as {\it type A}. We then give the conditions of type A
$\cN$-fold supersymmetry for arbitrary $\cN$. In the case of $\cN =2$,
the results reduces to just the ones reported in
Refs.\cite{And2,And3,And4}.
Section \ref{sec:speci} is devoted to illustrations of special cases
of type A, including the quartic and periodic potential cases.
In section \ref{sec:inter}, we reexamine the factorized intertwining
approach previously done in Ref.\cite{And2} and compare the results
with those in section \ref{sec:typeA}. We will see that novel
intermediate relations, which were not considered in Ref.\cite{And2}
at all, are allowed in order to get the $\cN$-fold superpartner.
Concluding remarks are in the last section.

\section{Review of the $\cN$-fold Supersymmetry}
\label{sec:nsusy}
First of all, we review the $\cN$-fold supersymmetry in one-dimensional
quantum mechanics \cite{And1,And2,Sam1,Sam2,Aoy1,Aoy2}
including the ordinary supersymmetric case \cite{Wit1,Wit2,Sol1,Coo1}.
The $\cN$-fold supercharges are generally defined in matrix form
by the following;
\begin{eqnarray}
Q_{\cN}=\left(
 \begin{array}{cc}
 0 & 0\\
 P_{\cN}^{\dagger}& 0
 \end{array}
\right), \quad Q_{\cN}^{\dagger}=\left(
 \begin{array}{cc}
 0 & P_{\cN}\\
 0 & 0
 \end{array}
\right), 
\label{eqn:nscdef}
\end{eqnarray}
where $P_{\cN}$ is a differential operator of order $\cN$.
The general form of $P_{\cN}$ is thus given by
\begin{eqnarray}
P_{\cN}=p^{\cN}+w_{\cN -1}(q)p^{\cN -1}+\cdots +w_{1}(q)p+w_{0}(q).
\label{eqn:nscge}
\end{eqnarray}
where $p=-i(d/dq)$. Clearly $Q_{\cN}$ and $Q_{\cN}^{\dagger}$ are
nilpotent, or equivalently,
\begin{eqnarray}
\{Q_{\cN}, Q_{\cN}\}=\{Q_{\cN}^{\dagger}, Q_{\cN}^{\dagger}\}=0.
\label{eqn:nsccom}
\end{eqnarray}
The Hamiltonian ${\bf H}_{\cN}$ defined in matrix form as
\begin{eqnarray}
{\bf H}_{\cN}=\left(
 \begin{array}{cc}
 H_{+\cN} & 0\\
 0 & H_{-\cN}
 \end{array}
\right),
\label{eqn:nsham}
\end{eqnarray}
is said to be $\cN$-fold supersymmetric if it commutes with
the $\cN$-fold supercharges;
\begin{eqnarray}
[Q_{\cN}, {\bf H}_{\cN}]=[Q_{\cN}^{\dagger}, {\bf H}_{\cN}]=0.
\label{eqn:nshcom}
\end{eqnarray}
The components of the above relations (\ref{eqn:nshcom}) are
\begin{eqnarray}
P_{\cN}H_{-\cN}-H_{+\cN}P_{\cN}=0
\label{eqn:nshrel}
\end{eqnarray}
and its hermitian conjugate.
The anticommutator of $Q_{\cN}^{\dagger}$ and $Q_{\cN}$ now becomes
a differential operator of order $2\cN$. Therefore, if the component
Hamiltonians of ${\bf H}_{\cN}$ are given by the following ordinary
Schr{\"o}dinger type;
\begin{eqnarray}
H_{\pm\cN}=
\frac{1}{2}\left(p^2+W(q)^2+V_{\pm\cN}(q)\right),
\label{eqn:nshpm}
\end{eqnarray}
the anticommutator can be generally expressed by a $\cN$-th
order polynomial $\cP_{\cN}$ of the Hamiltonian ${\bf H}_{\cN}$;
\begin{eqnarray}
\cH_{\cN}\equiv \frac{1}{2}\{Q_{\cN}^{\dagger}, Q_{\cN}\}
=\cP_{\cN}({\bf H}_{\cN}).
\label{eqn:motham}
\end{eqnarray}
The operator $\cH_{\cN}$ defined above is called the
{\it Mother Hamiltonian\/} and satisfies the following commutation
relations;
\begin{eqnarray}
[Q_{\cN}, \cH_{\cN}]=[Q_{\cN}^{\dagger}, \cH_{\cN}]=0.
\label{eqn:mhcom}
\end{eqnarray}

In the case of $\cN=1$, the $\cN$-fold supersymmetry defined above
reduces to the ordinary supersymmetric quantum
mechanics \cite{Wit1,Wit2,Sol1,Coo1}. Explicitly if we put,
\begin{eqnarray}
P_{1}\equiv D=p-iW(q), \quad
P_{1}^{\dagger}\equiv D^{\dagger}=p+iW(q),
\label{eqn:oscdef}
\end{eqnarray}
we immediately get the ordinary superalgebra;
\begin{mathletters}
\label{eqns:oscalg}
\begin{eqnarray}
&&\{ Q_{1}, Q_{1}\}=\{ Q_{1}^{\dagger}, Q_{1}^{\dagger}\} =0,\\
&&\{ Q_{1}^{\dagger}, Q_{1}\}=2{\bf H}_{1},\\
&&[Q_{1}, {\bf H}_{1}]=[Q_{1}^{\dagger}, {\bf H}_{1}]=0.
\end{eqnarray}
\end{mathletters}
The component Hamiltonians of ${\bf H}_{1}$ are given by
\begin{eqnarray}
H_{\pm 1}=\frac{1}{2}\left(p^2+W^2(q)\pm W'(q)\right).
\label{eqn:oshpm}
\end{eqnarray}
Comparing with the expressions (\ref{eqn:nshpm}) and
(\ref{eqn:motham}) we yield the relations;
\begin{eqnarray}
V_{\pm 1}(q)=\pm W'(q),
\quad\cH_{1}=\cP_{1}({\bf H}_{1})={\bf H}_{1}.
\label{eqn:osprel}
\end{eqnarray}

\section{type A $\cN$-fold Supersymmetry}
\label{sec:typeA}
In the previous paper \cite{Aoy2}, it was proved that if
the $\cN$-fold supercharges are limited to the form;
\begin{eqnarray}
P_{\cN}=D^{\cN}, \quad D=p-iW(q),
\label{eqn:quansc}
\end{eqnarray}
the $\cN$-fold supersymmetry can be realized only for quadratic $W(q)$.
It was also shown that for a periodic $W(q)$ with periodicity
$2\pi /g$, the system can possess $\cN$-fold supersymmetry with
respect to the following form of the $\cN$-fold supercharge;
\begin{eqnarray}
P_{\cN}=\prod_{k=-(\cN -1)/2}^{(\cN -1)/2}(D+kg).
\label{eqn:pernsc}
\end{eqnarray}
These facts indicate that the allowed $\cN$-fold supersymmetric
systems are characterized and limited by the form of the $\cN$-fold
supercharges. Motivated by this observation, we investigate a
particular class of the $\cN$-fold supercharges, which is called
{\it type A}. The form of the type A $\cN$-fold supercharges
$P_{\cN}^{(A)}$ is defined as follows;
\begin{eqnarray}
P_{\cN}^{(A)}&=&\Bigl(D+i(\cN -1)E(q)\Bigr)
 \Bigl(D+i(\cN -2)E(q)\Bigr)\cdots\Bigl(D+iE(q)\Bigr)D\nonumber\\
&\equiv&\prod_{k=0}^{\cN -1}\Bigl(D+ikE(q)\Bigr).
\label{eqn:tyInsc}
\end{eqnarray}
We will prove that the conditions of the Hamiltonian (\ref{eqn:nsham})
with (\ref{eqn:nshpm}) to be type A $\cN$-fold supersymmetric, that is,
to satisfy the relation (\ref{eqn:nshrel}), are as the following;
\begin{mathletters}
\label{eqns:nfscon}
\begin{eqnarray}
V_{\pm\cN}(q)&=&
 -(\cN -1)E(q)W(q)+\frac{(\cN -1)(2\cN -1)}{6}E(q)^2\nonumber\\
 &&{}-\frac{\cN^{\, 2}-1}{6}E'(q)\pm\cN\left(W'(q)-\frac{\cN -1}{2}E'(q)
 \right),
\label{eqn:nfsco1}\\
W(q)&=&\frac{E(q)}{2}+Ce^{-\int\! dqE(q)}\int\! dq\left(
 e^{\int\! dqE(q)}\int\! dq e^{\int\! dqE(q)}\right) \quad (\cN\geq 2),
\label{eqn:nfsco2}\\
E'''(q)&+&E(q)E''(q)+2E'(q)^2-2E(q)^2 E'(q)=0 \quad (\cN\geq 3).
\label{eqn:nfsco3}
\end{eqnarray}
\end{mathletters}
We can prove the above conditions
(\ref{eqns:nfscon}) 
by induction.
For $\cN =1$, the above
(\ref{eqns:nfscon}) 
(actually, only the
Eq.(\ref{eqn:nfsco1}) is applied) reads $V_{\pm1}(q)=\pm W'(q)$,
which is the ordinary supersymmetric case.

Suppose the relation (\ref{eqn:nshrel}) holds for an integer $\cN$.
Then, if we put
\begin{eqnarray}
H_{\pm (\cN +1)}=H_{\pm\cN}\pm h_{\pm\cN},
\label{eqn:diffhn}
\end{eqnarray}
and use the relation (\ref{eqn:nshrel}) for this $\cN$, we obtain,
\begin{eqnarray}
P_{\cN +1}^{(A)}H_{-(\cN +1)}-H_{+(\cN +1)}P_{\cN +1}^{(A)}
=[D+i\cN E, H_{+\cN}]P_{\cN}^{(A)}-h_{+\cN}P_{\cN +1}^{(A)}
 -P_{\cN +1}^{(A)}h_{-\cN}.
\label{eqn:diff1}
\end{eqnarray}
To facilitate the calculation, we make use of a similarity
transformation by $U$, which is defined by
\begin{eqnarray}
U=e^{\int\! dqW(q)}.
\label{eqn:defU}
\end{eqnarray}
The transformation of (\ref{eqn:diff1}) is then calculated as
\begin{eqnarray}
I_{\cN +1}&\equiv&2i^{\cN +1}U(P_{\cN +1}^{(A)}H_{-(\cN +1)}
 -H_{+(\cN +1)}P_{\cN +1}^{(A)})U^{-1}\nonumber\\
&=&[\partial -\cN E, -\partial^2 +2W\partial +W'+V_{+\cN}]\tP_{\cN}^{(A)}
 -2h_{+\cN}\tP_{\cN +1}^{(A)}-2\tP_{\cN +1}^{(A)}h_{-\cN}\nonumber\\
&=&2(W'-\cN E'-h_{+\cN}-h_{-\cN})\partial\tP_{\cN}^{(A)}\nonumber\\
&&{}+\Bigl(V'_{+\cN}+W''-\cN E''+2\cN E'W+2\cN (h_{+\cN}+h_{-\cN})E\Bigr)
 \tP_{\cN}^{(A)}
-2[\tP_{\cN +1}^{(A)}, h_{-\cN}],
\label{eqn:diff2}
\end{eqnarray}
where
\begin{eqnarray}
\tP_{\cN}^{(A)}&\equiv&i^{\cN}UP_{\cN}^{(A)}U^{-1}\nonumber\\
&=&\Bigl(\partial -(\cN -1)E(q)\Bigr)\Bigl(\partial -(\cN -2)E(q)\Bigr)
 \quad\cdots\quad \Bigl(\partial -E(q)\Bigr)\partial\nonumber\\
&\equiv&\prod_{k=0}^{\cN -1}\Bigl(\partial -kE(q)\Bigr).
\label{eqn:ptilde}
\end{eqnarray}
From Eq.(\ref{eqn:diff2}), we see that $I_{\cN +1}$ contains up to
$(\cN +1)$-th derivative. Therefore, $I_{\cN +1}=0$ if and only if
all the coefficients of $\partial^{k}$ ($k=0,1,\dots,\cN +1$) vanish.
The $\partial^{\cN +1}$ term comes only from the first term of the
r.h.s. of (\ref{eqn:diff2}) and thus,
\begin{eqnarray}
h_{+\cN}+h_{-\cN}=W'-\cN E'.
\label{eqn:del1}
\end{eqnarray}
When this condition (\ref{eqn:del1}) satisfied, the difference
$I_{\cN +1}$ now reads
\begin{eqnarray}
I_{\cN +1}&=&(V_{+\cN}'+W''-\cN E''+2\cN E'W+2\cN EW'-2\cN^{\, 2} EE')
 \tP_{\cN}^{(A)}
-2[\tP_{\cN +1}^{(A)}, h_{-\cN}].
\label{eqn:diff3}
\end{eqnarray}
The second term of the r.h.s. of (\ref{eqn:diff3}) is calculated
as follows;
\begin{eqnarray}
[\tP_{\cN +1}^{(A)}, h_{-\cN}]
&=&h_{-\cN}'\tP_{\cN}^{(A)}
 +(\partial -\cN E)[\tP_{\cN}^{(A)}, h_{-\cN}]\nonumber\\
&=&h_{-\cN}'\tP_{\cN}^{(A)}+(\partial -\cN E)
 \Biggl[\cN h_{-\cN}'\partial^{\cN -1}\nonumber\\
&&{}+\frac{\cN (\cN -1)}{2}\Bigl(h_{-\cN}''
 -(\cN -1)Eh_{-\cN}'\Bigr)\partial^{\cN -2}+\cdots\Biggr],
\label{eqn:oofdef}
\end{eqnarray}
where $\cdots$ denotes the terms which contain up to the ($\cN -2$)-th
order derivative. From the $\partial^{\cN}$ and $\partial^{\cN -1}$
terms, we obtain the following conditions respectively;
\begin{eqnarray}
2(\cN +1)h_{-\cN}'&=&V_{+\cN}'+W''-\cN E''
+2\cN E'W+2\cN EW'-2\cN^{\, 2}EE',
\label{eqn:del2}
\end{eqnarray}
\begin{eqnarray}
h_{-\cN}''-Eh_{-\cN}'=0.
\label{eqn:del3}
\end{eqnarray}
The condition (\ref{eqn:del2}) can be easily integrated, and with
the condition (\ref{eqn:del1}) we get
\begin{eqnarray}
\pm h_{\pm\cN}=\frac{1}{2}\left[-EW+\frac{4\cN -1}{6}E^{2}
 -\frac{2\cN +1}{6}E'\pm (W'-\cN E')\right].
\label{eqn:pmhpmn}
\end{eqnarray}
Here we omit the irrelevant integral constants.
Therefore, we finally yield
\begin{eqnarray}
V_{\pm (\cN +1)}&=&V_{\pm\cN}\pm 2h_{\pm\cN}\nonumber\\
&=&-\cN EW+\frac{\cN (2\cN +1)}{6}E^{2}
-\frac{\cN (\cN +2)}{6}E'
 \pm (\cN +1)\left(W'-\frac{\cN}{2}E'\right),
\label{eqn:vpmn+1}
\end{eqnarray}
which are nothing but the assumed forms of the potential
(\ref{eqn:nfsco1}) with $\cN$ replaced by $\cN +1$.
Before investigating the condition (\ref{eqn:del3}), we return to
the difference $I_{\cN +1}$ under the condition (\ref{eqn:del2}),
which now reads
\begin{eqnarray}
I_{\cN +1}=2\cN h_{-\cN}'\tP_{\cN}^{(A)}-2(\partial -\cN E)
 [\tP_{\cN}^{(A)}, h_{-\cN}].
\label{eqn:diff4}
\end{eqnarray}
It is easy to see that under the condition (\ref{eqn:del3}), the
following relation holds;
\begin{eqnarray}
[\tP_{\cN}^{(A)}, h_{-\cN}]=Mh_{-\cN}'\tP_{\cN -1}^{(A)}
 +[\prod_{k=M}^{\cN -1}(\partial -kE), h_{-\cN}]\tP_{M}^{(A)}
 \quad (0\leq M\leq \cN).
\label{eqn:comtp}
\end{eqnarray}
Applying this relation (with $M=\cN$) to Eq.(\ref{eqn:diff4}), we
immediately find $I_{\cN +1}=0$. That is, no additional
conditions are needed for satisfying the relation (\ref{eqn:nshrel})
with $\cN +1$. So, all that remains to be investigated is the condition
(\ref{eqn:del3}). From Eq.(\ref{eqn:pmhpmn}), this condition reads,
\begin{eqnarray}
(W_{\cN}'+EW_{\cN})''-E(W_{\cN}'+EW_{\cN})'=0,
\label{eqn:confe1}
\end{eqnarray}
where
\begin{eqnarray}
W_{\cN}(q)=W(q)-\frac{4\cN -1}{6}E(q).
\label{eqn:defwn}
\end{eqnarray}
In the case of $\cN =1$, the condition (\ref{eqn:confe1}) gives
the relation between $W(q)$ and $E(q)$;
\begin{eqnarray}
\left[\left(W-\frac{E}{2}\right)'
 +E\left(W-\frac{E}{2}\right)\right]''
-E\left[\left(W-\frac{E}{2}\right)'
 +E\left(W-\frac{E}{2}\right)\right]'=0.
\label{eqn:n1cond}
\end{eqnarray}
Equation (\ref{eqn:n1cond}) can be
integrated for $W(q)$ in terms of $E(q)$, which leads to
the condition (\ref{eqn:nfsco2}).
In the case of $\cN\geq 2$, the condition (\ref{eqn:confe1}) should be
compatible with that for $\cN =1$ by the inductive assumption.
This immediately leads to
\begin{eqnarray}
(E'+E^2)''-E(E'+E^2)'=0,
\label{eqn:n2cond}
\end{eqnarray}
which is equivalent to the last condition (\ref{eqn:nfsco3})
and the proof is completed.

It is tempting from the potential form (\ref{eqn:nfsco1}) to
redefine the prepotential as
\begin{eqnarray}
\tW(q)\equiv W(q)-\frac{\cN -1}{2}E(q).
\label{eqn:deftw}
\end{eqnarray}
From the conditions (\ref{eqn:n1cond}) and (\ref{eqn:n2cond}),
$\tW$ should satisfy
\begin{eqnarray}
(\tW'+E\tW )''-E(\tW'+E\tW )'=0 \quad\textrm{for}\quad\cN\geq 2.
\label{eqn:confe2}
\end{eqnarray}
With this $\tW(q)$, we obtain another general form of type A $\cN$-fold
supersymmetry;
\begin{mathletters}
\label{eqns:anfscon}
\begin{eqnarray}
&&P_{\cN}^{(A)}=\prod_{k=-(\cN -1)/2}^{(\cN -1)/2}
 \Bigl(\tD +ikE(q)\Bigr),\quad \tD=p-i\tW(q),
\label{eqn:anfsco0}\\
&&2H_{\pm\cN}=p^2 +\tW(q)^2 +\frac{\cN^{\, 2}-1}{12}
 \Bigl(E(q)^2-2E'(q)\Bigr)\pm\cN \tW'(q),
\label{eqn:anfsco1}\\
&&\tW(q)=Ce^{-\int\! dqE(q)}\int\! dq\left(
 e^{\int\! dqE(q)}\int\! dq\, e^{\int\! dqE(q)}\right) \quad (\cN\geq 2),
\label{eqn:anfsco2}\\
&&E'''(q)+E(q)E''(q)+2E'(q)^2-2E(q)^2 E'(q)=0 \quad (\cN\geq 3).
\label{eqn:anfsco3}
\end{eqnarray}
\end{mathletters}
Furthermore, we can express the Hamiltonians (\ref{eqn:anfsco1}) solely
in terms of the prepotential $\tW (q)$. From the condition
(\ref{eqn:confe2}), an useful relation holds;
\begin{eqnarray}
\bigl[\tW^2 (E^2 -2E')\bigr]'=2\tW\tW'''.
\label{eqn:relwae}
\end{eqnarray}
Using this equality we yield, instead of Eq.(\ref{eqn:anfsco1}),
\begin{eqnarray}
2H_{\pm\cN}&=&p^2 +\tW (q)^2
+\frac{\cN^{\, 2}-1}{12}\left(
 \frac{2\tW''(q)}{\tW (q)}-\frac{\tW'(q)^2}{\tW (q)^2}
 +\frac{A}{\tW (q)^2}\right)\pm\cN\tW'(q),
\label{eqn:shambw}
\end{eqnarray}
where $A$ is an arbitrary constant. In the case of $\cN =2$, the
above (\ref{eqn:shambw}) is reduced to the result obtained in
Ref.\cite{And2} for the second order Darboux transformation.

\section{Special Cases of Type A}
\label{sec:speci}
In this section, we illustrate some special cases of the type A
$\cN$-fold supersymmetry by using the general results obtained in
the previous section. We will see that the quadratic and the periodic
$W(q)$s which were earlier found to possess the $\cN$-fold
supersymmetry \cite{Aoy1,Aoy2} can be obtained in this way.

First of all, we set $E(q)=0$. This is a trivial solution of
Eq.(\ref{eqn:nfsco3}). From Eq.(\ref{eqn:nfsco2}) we yield,
\begin{eqnarray}
W(q)=C_{1}q^2 +C_{2}q +C_{3},
\label{eqn:quadrw}
\end{eqnarray}
that is, quadratic $W(q)$. In this case, the Hamiltonians and
the supercharge are given by
\begin{eqnarray}
2H_{\pm\cN}=p^2 +W(q)^2 \pm\cN W'(q),\quad P_{\cN}^{(A)}=D^{\cN}.
\label{eqn:spham1}
\end{eqnarray}
The special choices $C_{1}=-g$, $C_{2}=1$ and $C_{3}=0$ correspond
to just the case in Ref.\cite{Aoy1}.

In the next, we set $E(q)=E_{0}$(non-zero constant). This is also
a trivial solution of Eq.(\ref{eqn:anfsco3}). From
Eq.(\ref{eqn:anfsco2}) we yield,
\begin{eqnarray}
W(q)=C_{1}e^{E_{0}q}+C_{2}e^{-E_{0}q}+C_{3},
\label{eqn:periow}
\end{eqnarray}
that is, exponential $W(q)$. In this case, the Hamiltonians and
the supercharge are given by
\begin{eqnarray}
2H_{\pm\cN}=p^2 +W(q)^2 \pm\cN W'(q),\quad
 P_{\cN}^{(A)}=\prod_{k=-(\cN -1)/2}^{(\cN -1)/2}\Bigl(D +ikE_{0}\Bigr).
\label{eqn:spham2}
\end{eqnarray}
The special choices $E_{0}=ig$, $C_{1}=1/2ig$, $C_{2}=-1/2ig$ and
$C_{3}=0$ correspond to the periodic case in Ref.\cite{Aoy2}.

Next, we set $E(q)=(\nu -1)/q$ with $\nu\ne 1$. It is easy to see
that Eq.(\ref{eqn:anfsco3}) is satisfied when $\nu =\pm 2$.
In both the cases we get from Eq.(\ref{eqn:anfsco2}),
\begin{eqnarray}
W(q)=C_{1}q^3 +C_{2}q+C_{3}\frac{1}{q}.
\label{eqn:cubicw}
\end{eqnarray}
In these cases, the Hamiltonians are
\begin{eqnarray}
2H_{\pm\cN}=p^2 +W(q)^2 +\frac{\cN^{\, 2}-1}{4q^2}\pm\cN W'(q),
\label{eqn:spham3}
\end{eqnarray}
and the supercharges are given by
\begin{eqnarray}
P_{\cN}^{(A)}=\prod_{k=0}^{N-1}\left(D+i\frac{k}{q}\right)
\quad (\nu =+2),
\quad\textrm{or}\quad
P_{\cN}^{(A)}=\prod_{k=0}^{N-1}\left(D-3i\frac{k}{q}\right)
\quad (\nu =-2).
\label{eqn:spchar}
\end{eqnarray}
This cubic type $W(q)$ is a new form of the $\cN$-fold supersymmetry.
It should be noted that the $\cN$-fold supercharges for one
Hamiltonian pair permits different factorized forms in general,
as is the case above (\ref{eqn:spchar}), owing to the fact that $E(q)$
satisfies a differential equation (\ref{eqn:nfsco3}).

\section{Factorized Intertwining Approach}
\label{sec:inter}
We note the type A $\cN$-fold supercharges belong to {\it reducible\/}
$\cN$-th order intertwiners $L_{\cN}$, which can be factorized as
a product of $\cN$ first order differential operators $L^{(k)}$
\cite{And2,And3,And4};
\begin{eqnarray}
L_{\cN}=L^{(\cN)}\cdots L^{(1)}.
\label{eqn:inter}
\end{eqnarray}
For such a reducible operator, the factorized intertwining
technique\cite{And2} can be applicable. In this approach,
a $\cN$-fold supersymmetric model is constructed by introducing
a sequence of intermediate Hamiltonians $H^{(k)}$, which satisfy
the ordinary supersymmetric relations;
\begin{eqnarray}
H^{(k)}L^{(k)}=L^{(k)}H^{(k-1)}\quad (k=1, 2, \dots, \cN).
\label{eqn:intham}
\end{eqnarray}
Apparently, the following $\cN$-fold supersymmetric relation
\begin{eqnarray}
H_{+\cN}L_{\cN}=L_{\cN}H_{-\cN},
\label{eqn:intnfs}
\end{eqnarray}
holds if we set,
\begin{eqnarray}
H_{+\cN}=H^{(\cN)},\quad H_{-\cN}=H^{(0)}.
\label{eqn:defphm}
\end{eqnarray}
In this section we reexamine
the conditions of type A $\cN$-fold supersymmetry by this
intertwining approach and compare the results with those obtained in
section \ref{sec:typeA}. The type A $\cN$-fold supercharge is realized
if we set the each factor of a intertwiner as
\begin{eqnarray}
L^{(k)}\equiv D+i(k-1)E(q)=p-i\Bigl(W(q)-(k-1)E(q)\Bigr).
\label{eqn:1stint}
\end{eqnarray}
Each of the above $L^{(k)}$ can be regarded as an ordinary supercharge
with prepotential $W-(k-1)E$. Therefore, if we introduce
Hamiltonians $H_{>}^{(k)}$ and $H_{<}^{(k)}$ as,
\begin{mathletters}
\label{eqns:inpmham}
\begin{eqnarray}
2H_{>}^{(k)}&=&L^{(k)}L^{(k)\dagger}+2C(k)\nonumber\\
 &=&p^2 +\Bigl(W-(k-1)E\Bigr)^2 +\Bigl(W-(k-1)E\Bigr)'+2C(k),\\
2H_{<}^{(k-1)}&=&L^{(k)\dagger}L^{(k)}+2C(k)\nonumber\\
 &=&p^2 +\Bigl(W-(k-1)E\Bigr)^2 -\Bigl(W-(k-1)E\Bigr)'+2C(k),
\end{eqnarray}
\end{mathletters}
where $C(k)$s are arbitrary constants,
these Hamiltonians satisfy the supersymmetric relation for each $k$;
\begin{eqnarray}
H_{>}^{(k)}L^{(k)}=L^{(k)}H_{<}^{(k-1)}.
\label{eachsr}
\end{eqnarray}
For the above supersymmetric Hamiltonians constructed in each $k$
together to construct the $\cN$-fold supersymmetry, the following
conditions should be satisfied;
\begin{eqnarray}
H_{>}^{(k)}=H_{<}^{(k)}\quad (k=1, \dots , \cN -1).
\label{eqn:necinh}
\end{eqnarray}
This kind of intermediate
relations were actually considered in Ref.\cite{And2}.
Explicitly, this condition is expressed as
\begin{eqnarray}
\left(W-\frac{E}{2}\right)'+E\left(W-\frac{E}{2}\right)-c_{1}
=(k-1)\Bigl(E'+E^2-c(k)\Bigr)\ (k=1, \dots , \cN -1),
\label{eqn:neccmp}
\end{eqnarray}
where we put $C(k+1)-C(k)=c_{1}-(k-1)c(k)$.
For $\cN =2$, the above condition (\ref{eqn:neccmp}) reads
\begin{eqnarray}
\left(W-\frac{E}{2}\right)'+E\left(W-\frac{E}{2}\right)=c_{1}.
\label{eqn:infcon1}
\end{eqnarray}
For $\cN\geq 3$, to fulfill Eq.(\ref{eqn:neccmp}) for arbitrary $k$,
$c(k)$ should not depend on $k$ and thus we put $c(k)\equiv c$,
and the following is needed,
\begin{eqnarray}
E'+E^2 =c,
\label{eqn:infcon2}
\end{eqnarray}
in addition to Eq.(\ref{eqn:infcon1}).
Comparing these results (\ref{eqn:infcon1}) and (\ref{eqn:infcon2})
with the conditions obtained in section \ref{sec:typeA}, we see that
the results (\ref{eqn:infcon1}) and (\ref{eqn:infcon2}) are sufficient
conditions for satisfying Eq.(\ref{eqn:n1cond}) and
Eq.(\ref{eqn:n2cond}), respectively.
Conversely, $c_{1}$
and $c$ are not necessarily constant but can be functions of $q$,
which satisfy,
\begin{eqnarray}
c_{(1)}''(q)-E(q)\, c_{(1)}'(q)=0.
\label{eqn:difeqc}
\end{eqnarray}
This result indicates that even in the reducible cases there may be
wider class of $\cN$-fold supersymmetric models than that can be
obtained by the factorized intertwining technique.

\section{Concluding Remarks}
\label{sec:concl}
In this letter, we have shown the general forms and conditions of a
$\cN$-fold supersymmetric family. Using the results, one
can easily obtain a $\cN$-fold supersymmetric model for arbitrary
$\cN$ with or without dynamical SUSY breaking. If dynamical
SUSY breaking takes place or not depends on the domain in which
the system is defined and on the asymptotic behavior of $W(q)$.
Though the specific type investigated in this paper is
quite general, It will be an interesting problem to find another
type of family which does not belong to type A.

Finally, we will mention about the non-renormalization theorem.
This theorem is one of the most notable properties that the
supersymmetric models possess. However, little has been discussed
about the theorem in the case of the $\cN$-fold supersymmetry.
As far as we know, only Ref.\cite{Aoy1} investigated
the non-renormalization nature for the quartic $W(q)$ case.
We have found the same property for the
other $\cN$-fold supersymmetric models such as the periodic and
the cubic $W(q)$s illustrated in section \ref{sec:speci}.
These results will be reported in the near future.

\acknowledgments

The authors would like to thank Dr.~Hisashi Kikuchi
(Ohu University, Japan) for discussions.
The authors would also like to thank Dr.~V.~P.~Spiridonov and
Dr.~M.~Plyushchay for information on their works.
H.~Aoyama's work was supported in part by the Grant-in-Aid 
for Scientific Research No.10640259.
T.~Tanaka's work was supported in part by a JSPS research fellowship.




\end{document}